# Development of a Large-scale Neuroimages and Clinical Variables Data Atlas in the neuGRID4You (N4U) project

**Kamran Munir***, **Khawar Hasham Ahmad, Richard McClatchey**

*Centre for Complex Cooperative Systems (CCCS), Department of Computer Science and Creative Technologies (CSCT), University of the West of England (UWE), Frenchay Campus, Coldharbour Lane, Bristol, BS16 1QY, United Kingdom.*

**Abstract.**

*Exceptional growth in the availability of large-scale clinical imaging datasets has led to the development of computational infrastructures offering scientists access to image repositories and associated clinical variables' data. The EU FP7 neuGRID and its follow on neuGRID4You (N4U) project is a leading e-Infrastructure where neuroscientists can find core services and resources for brain image analysis. The core component of this e-Infrastructure is the N4U Virtual Laboratory, which offers an easy access for neuroscientists to a wide range of datasets and algorithms, pipelines, computational resources, services, and associated support services. The foundation of this virtual laboratory is a massive data store plus information services called the 'Data Atlas' that stores datasets, clinical study data, data dictionaries, algorithm/pipeline definitions, and provides interfaces for parameterised querying so that neuroscientists can perform analyses on required datasets. This paper presents the overall design and development of the Data Atlas, its associated datasets and indexing and a set of retrieval services that originated from the development of the N4U Virtual Laboratory in the EU FP7 N4U project in the light of user requirements.*

## 1. Introduction

With the enormous increase in variety and size of clinical datasets, and increasing information complexity in biomedical research, biomedical researchers are often faced with severe difficulties in data management and information retrieval. The neuGRID4You project (N4U Project, grant agreement n. 283562, 2011-2014) provides an e-infrastructure where neuroscientists can find services and resources for neuroimaging analysis [1]. It provides an e-Science environment through a Virtual Laboratory (whose model is depicted in Figure 1) that offers neuroscientists access to a wide range of clinical and neuroimaging datasets, algorithm applications, computational resources, services, and support. This virtual laboratory has been mainly developed for neuroscientists but is adaptable to other user communities. The foundation of this N4U Virtual Laboratory is a massive data store, named the Analysis Base, and Information Services. The combination of the Analysis Base and its Information Services is termed the 'Data Atlas'. The Analysis Base is a structured data store that stores datasets, clinical study data, data dictionaries, and algorithm/pipeline definitions. The Analysis Base is accessed through Information Services that store or index datasets and algorithm pipeline definitions required so that users can perform any investigation or data analysis through the N4U Virtual Laboratory. This subset of Information Services is termed the Persistency Services. Moreover, the Information Services also provide various interfaces for the parameterised querying of datasets to assist the virtual laboratory users (such as neuroscientists) in defining and executing their analyses on filtered datasets. This subset of Information Services is termed the Querying Services. The outcome generated by the Querying Service can be exported in various formats, such as XML and CSV, which is then used in other software applications to generate/perform an analysis, e.g. by using the CRISTAL software [2].

---

* This is the corresponding author
*Email address:* kamran2.munir@uwe.ac.uk (Kamran Munir)

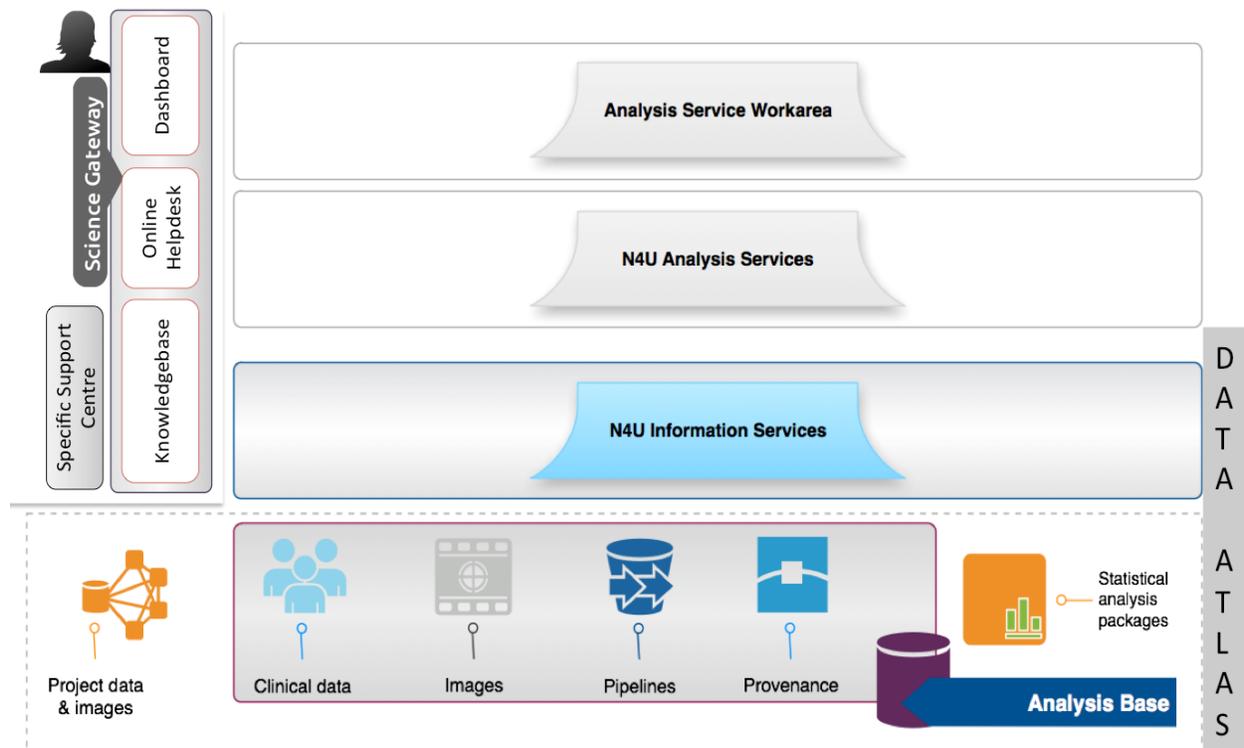

Figure 1: The N4U Virtual Laboratory architecture.

The design and development of the Analysis Base and Information Services (a.k.a. both persistency and querying services) has been carried out in the light of detailed N4U users' requirements [3], which have been analysed during the requirements gathering and specification phases of the N4U project. Moreover, there have been an enormous number of evolving requirements, such as the increased complexity and heterogeneity of datasets, which has led to further refinements in the design specifications and implementation of the Analysis Base and Information Services. The main challenges faced in providing these services relate to, for example: (a) the different formats, structures and semantics of datasets; (b) the fact that all datasets had a large diversity in the number and types of clinical study parameters; (c) the manner in which the original relationships between clinical study parameters and image files (such as brain scans) were maintained also varied across datasets and dataset providers; (d) the fact that comprehensive data dictionaries had to be maintained for each dataset; and (e) that each dataset can have unlimited numbers of clinical variables that can hold different types of values (called clinical variables' scores). This paper presents the overall design and development of the Data Atlas, including the development methodologies and algorithms used by its associated information services in the light of user requirements.

This paper has been organised as follows: Section 2 reviews the relevant approaches in neuroinformatics that aid in managing biomedical data and assist researchers in performing data analysis. In Section 3, the requirements, design and development of the N4U Analysis Base, which is the foundation of the N4U Virtual Laboratory, are presented. Section 4 outlines the requirements and associated challenges of dealing with large datasets, followed by an introduction to the N4U Information Services that carry out all necessary functions of importing, indexing and querying of the datasets in the Analysis Base including, for example, clinical study data, neuroimages, pipelines and algorithms. Section 5 presents the indexing of use cases (examples) of datasets, pipelines and algorithms, which show the availability of different types of datasets in N4U and also their internal structures – representing the medical data along with data dictionaries and image files. Section 6 discusses the implemented data retrieval mechanisms, which enable the N4U end users to perform their queries on clinical data and images in order to retrieve desired data related to pipelines/algorithms, for the exploration of the datasets in the Analysis Base. This section is concluded by a description of the use of rich clinical data dictionaries to empower end-users in defining their search criteria. Finally, outcomes and identified future research challenges are presented in Section 7.

## 2. Related Work

Managing huge volumes of data produced in neuroscience research and enabling their querying have been among the major research challenges in the neuroscience community [4]. There have been much work carried out to support the storing and retrieval of neuroscience data. Such efforts can be divided into two main categories; (a) the data of neuro-science resources including papers and (b) the actual neuroscience experimental data including neuroimages and clinical variables. These two categories have been discussed in the following subsections. The current work focuses on providing a mechanism to enable the indexing and retrieval of neuroimages stored on the Grid infrastructure of N4U, which is an extension of the neuGRID project (http://www.neugrid.eu).

### (a) Data of Neuroscience Resources

In the first category, [5] discusses some projects that been developed to discover and integrate neuroscience resources to facilitate the neuro-science community. The primary aim of these projects is to provide a single platform for neuroscience researchers to search for the required data based on given domain specific keywords. These also act as the basis of a search engine for neuroscientists to discover relevant articles as well as databases to locate selected data. The Neuroscience Database Gateway (NDG) (http://ndg.sfn.org/) is a Web-based database of neuroscience databases. It provides a registry of neuroscience databases annotated with controlled keywords. It supports over 200 databases spanning different neuroscience subdomains such as neurophysiology SenseLab [6] and neuroimaging. The Neuroscience Information Framework (NIF) [7] provides a "one-stop-shop" for neuroscience researchers to access heterogeneous information resources. It does so by providing three registries: (i) a NIF resource registry, (ii) a NIF database mediator and (iii) a NIF document archiver. NIF is an ontology driven system and at its heart is NIFSTD (NIF Standard Ontology), which is constructed by integrating various other domain ontologies. DISCO [8] is a web based application developed for NIF to provide features such as resource integration and data searching powered by NIF backend architecture. It provides concept-driven querying support for data discovery.

The KIND (Knowledge-Based Integration of Neuroscience Data) approach [9] overcomes the problem of data integration of datasets that differ in formats. The data integration is performed over the biological study data that come from different sources such as NTRANS (neuro-transmission database) and CAPROT (calcium-binding protein databases). In order to perform the data integration, it uses domain knowledge that describes rules of the domain to bridge the gap between disparate data sources. Unlike the Data Atlas, the KIND does not provide a user-friendly front-end interface that would allow a user to build and refine its search queries. It then applies the deductive object-oriented language *F-Logic* to support complex data integration. Similarly, Entrez Neuron [10] provides a keyword-based search against a coherent repository described in the Web Ontology Language (OWL) (http://www.w3.org/2001/sw/wiki/OWL) ontologies such as NeuroDB and ModelDB (from SenseLab) and Subcellular Anatomy Ontology (SAO) [11]. These ontologies are then stored in the Oracle 11g database (http://www.oracle.come) that has built-in support for the OWL storage and querying. Unlike the Entrez Neuron, the Data Atlas does not support OWL based data sources at the moment because the data exported to Data Atlas is not represented in the OWL format. Furthermore, the querying interface of Entrez Neuron is very basic as compared to the Data Atlas, which provides a rich and dynamic querying building interface (discussed in Section 6) to a user.

The NeuroLOG [12] project proposed a federated approach to integrate neuroimage files, associated metadata and neuroscience semantic data distributed across different sites. A Data Management Layer (DML) has been devised to hide the underlying complexity of data stored in different formats at different sites. In doing so, an additional NeuroLOG database, structured according to defined ontologies, is deployed on each site to achieve complete autonomy of the sites and to facilitate existing data formats stored on the sites. The main focus in NeuroLOG is on the storage and semantic retrieval of the data stored across different sites, however this approach does not provide provenance information of the neuroimaging analyses conducted over the neuroimages. In comparison, the N4U Data Atlas provides support for data indexing and user-driven query building, and the provenance information of user analyses can be retrieved through the CRISTAL software.

The work presented in this paper is different from aforementioned efforts on various counts. Firstly, the Data Atlas is designed to provide data indexing and querying services to search and locate actual datasets (brain scans) images and their associated clinical variables and metadata. This information is crucial to support user analyses executed by the Analysis Services using CRISTAL (see [13]) on the N4U infrastructure. Therefore it is essential to provide a mechanism that enables researchers to discover and access neuroimages stored on the Grid infrastructure v.i.z. the N4U Grid infrastructure. Secondly, aforementioned projects dealt with heterogeneous neuroscience resources, however they have not explored the heterogeneity within the neuro-data collected at various neuroscience research labs or centers. As discussed in Section 4.1 of this paper, in N4U each dataset is different from the others, consequently providing a uniform indexing and then querying mechanism becomes a challenging task. Moreover, the aforementioned projects use ontologies or predefined keywords, which are generated as a result of domain knowledge described in the OWL ontologies, in order to provide keyword- or concept- based searches. However, we have presented a dynamic query-building tool that presents all the clinical variables and their metadata to a user and then allows the user to build her own query with dynamic values to obtain her desired result set. In a way, this gives more flexibility and therefore is more customizable. Furthermore, most of the existing work is based on semantics, which is not the focus of this paper. However, the Data Atlas could be extended in future by integrating a semantic layer to provide this feature.

### (b) Neuroscience Experimental Data

There has also been a lot of research to support the storing and querying of actual clinical study data. FBIRN [14] presents an extensible data management system for clinical neuroimaging studies. They have developed a distributed network infrastructure to support the creation of a federated database consisting of a large sample of neuroimaging datasets. FBIRN also incorporates the provenance information of an analysis. However, this provenance information does not provide an insight about the execution times of an analysis. Furthermore, the federated multi-site data organization approach requires the configuration and deployment of the proposed framework at each site, with a consequent increases in overall complexity. In comparison our proposed approach is comparatively simpler since it provides interfaces for external data sources that can be used to index external data according to a single schema, thus providing a uniform storage view to the users and external data sources. In N4U all the transformation and complexity is hidden from the external data sources and is handled internally by the Persistency Service (as detailed in Section 5). In doing so, a standard data format has been devised for data providers in order to import their data into the N4U Analysis Base.

The CNARI framework [15], used at the Human Neuroscience Laboratory at the University of Chicago, presents a database-driven architecture that combines databases for storing the fMRI data and workflows for analysis purposes. It proposes the use of relational databases instead of a traditional file-based approach for storing and querying fMRI data [16]. It creates separate databases for each experiment and several tables to store images or user-specific data, which requires a huge storage space. In order to execute the workflows, CNARI employs SWIFT as a workflow engine [17] and exploits its provenance tracking capabilities to maintain provenance information for reproducing an analysis. Unlike CNARI, our proposed schema stores references to the image files available on the N4U Grid infrastructure, thus making it storage efficient. Moreover, it also provides runtime provenance information of each analysis through CRISTAL.

The Human Imaging Database (HID) [18] is an open-source and extensible database schema implemented over relational databases Oracle (http://www.oracle.com) and PostgreSQL (http://postgresql.org). This system has been designed to operate in a federated environment in which each site has its own HID instance. The images and derived data, which are physically stored on the Grid, are linked back to HID located at the site that imported the data into the system. Since HID operates in a federated environment, it provides a data integration engine so that the data stored on all sites can be exposed and queried as a single database.

The Neuroinformatics Database (NiDB) [19], developed at Hartford Hospital for the Olin Neuropsychiatry Research Center, also provides an open-source database and a pipeline management system. It provides a platform to store and manipulate neuroimaging data and addresses challenges of data sharing identified by the International Neuroinformatics Coordinating Facility (INCF) Task Force on Neuroimaging Datasharing. It not only supports local storage and analysis of neuroimaging data but it also provides data sharing feature. It also provides a search mechanism for pipeline results with predefined search criteria for which a user can provide values. However, it does not provide a way for a user to build a dynamic query for the clinical variables present in a dataset. This is the feature that is supported by the Data Atlas (discussed in Section 6). Moreover, the NiDB can also support creation and execution of pipelines on local clusters, however it is not clear from the description that it can also support execution of those pipelines on other distributed environments such as the Grid. However, In N4U, all user pipelines are executed on the Grid infrastructure and the datasets are also stored on the Grid. A user can locate the dataset's Logical File Name (lfn) based on the search criteria he constructed on the querying interface of the Data Atlas. This information can then be passed on to the N4U Analysis Service through CRISTAL to execute user analysis on the selected datasets.

The Extensible Neuroimaging Archive Toolkit (XNAT) [20], developed by the Neuroinformatics Research Group at Washington University at St. Louis, aims at offering researchers an integrated environment for the archival, searching and sharing of neuroimaging datasets. It relies on an extensible XML schema to represent imaging and experimental data and supports a relational database backend. It is aimed at managing large amounts of data via a three-tier design infrastructure consisting of a client front end, the XNAT middleware and a data store. The data store is composed of a relational database and a file system on which images are stored. A small portion of the schema presented in this paper has a purpose similar to XNAT i.e., to store pointers to the files in the database. However, the data storage and querying mechanism of Data Atlas differ from the XNAT's approach. The XNAT relies upon hybrid storage i.e., XSDs and relational database to represent and store data. The XNAT generates relational databases from the XSDs, which are site dependent. This dual representation requires changes to the data model percolate through the XSDs and the database, and this often requires manual intervention. However, this is not the case with Data Atlas that uses relational database only. Moreover, the primary querying language in XNAT is based on the XPath (www.w3c.org/TR/xpath) and the XQuery (www.w3c.org/TR/xquery), however the Data Atlas uses SQL as a query language. Furthermore, the web forms in XNAT are also generated from the XSDs schemas in contrast to the dynamic form generation in the Data Atlas using the values from its database.

GridPACS [21] supports the distributed storage, retrieval and querying of image data and associated descriptive metadata. Workflows and metadata are modeled as XML schemas unlike the relational database schema approach used in the N4U analysis base, which supports rich querying using SQL. It integrates image data and metadata to maintain provenance information about how the images have been acquired and processed. SenseLab [6], developed at Yale University, is a metadata driven system to store scientific data using an entity-attribute-value with classes and relationships representation in a relational database. Most of these approaches primarily focus on the storage of clinical data; however, the N4U Analysis Base not only provides a schema to store and retrieve image datasets but also contains a layout for storing associated clinical study data with subjects linked to image datasets, and pipelines.

## 3.  The N4U Analysis Base

In N4U the actual datasets are stored on the N4U Grid-based infrastructure, however providing access to these datasets is not trivial, because a user, who may want to carry out an analysis, may need to select part of the dataset based on specific characteristics. Filtering gigabytes of data at runtime is a non-trivial task, unless it is methodically indexed beforehand. Furthermore, creating such indexes also becomes challenging when the metadata associated with the datasets is stored separately [22]. The challenge is further compounded by the presence of disparate formats of various datasets. The N4U Analysis Base has been designed in a way that overcomes these challenges and provides functionality in the N4U project such as: (i) access to indexed images and datasets [23]; (ii) index workflow/pipeline and algorithm definitions; (iii) provide interfaces for data export into the Analysis Base; and (iv) linkage between neuroimages and clinical data to provide an effective data retrieval facility through ad-hoc queries. In this way, the N4U Analysis Base offers an integrated medical data analysis environment to optimally exploit neuroscience pipelines, large image datasets and algorithms for clinical analyses. This interlinking of pipelines, algorithms, clinical variables and clinical images was one of the major aims of developing the Analysis Base in order to conduct neuroscience analyses and process complex users' queries. The support for user defined queries enables a user to identify the location(s) of external data stored on the Grid infrastructure. These queries involve (but are not limited to) (i) searches for datasets or image files locations, their creation dates and modification times if these files are modified over a period of time; and (ii) searches for a pipeline or algorithm(s) by name, or with specific search requirements such as those pipeline(s) or algorithm(s) created by a specific user, etc.

Moreover, a few of the important constraints that appeared during the course of discussions and which influenced the design of the Analysis Base database schema are: (a) a user can create none or multiple pipelines (a.k.a. workflows), a pipeline can have multiple algorithms, and a single algorithm can be used in multiple pipelines; and (b) a dataset, which is a collection of files, can have different combinations of image and data files. A dataset can further be organised into subcategories and assessments. The clinical data and image files will have a defined inter-relationship.

The layout of the Analysis Base schema illustrating the entities and their mutual relationships is shown as Figure 2. In this schema, the user information is linked to N4U pipelines, algorithms and datasets as stored in the Analysis Base. The *User* entity in the schema uniquely identifies a user and its associated role that is primarily maintained in order to respond to the queries related to the user roles. Any change in an user's information or registration is also managed. The *Group* entity defines a collection of users and helps in understanding their various possible roles such as researcher, practitioner, or administrator. Since each user can be associated with multiple groups and a group can have multiple users, an intermediary entity *User_Group* has been created to maintain this relationship. This *User_Grou*p provides the *many-to-many* relationship required between the *User* and *Group*. The *Pipeline* entity maintains the definition of a pipeline such as the pipeline *version* and the pipeline *description* in the N4U community. Each pipeline is uniquely identified in the Analysis Base and each user can be associated with *one-to-many* pipelines.

An *Algorithm* in N4U (also defined in the Analysis Base schema) is basically a set of tasks that a neuroscientist wants to perform over one or more neuro-datasets. Each algorithm is assigned a unique identifier. The algorithms' physical location is external to the Analysis Base; and therefore, the *Algorithm* entity maintains a *Logical File Name (lfn)* to address an actual algorithm stored on the N4U Grid infrastructure, which is translated to a physical location by the Replica Location Services (RLS) or catalogue [24]. It may be possible that some files are not stored in the Grid infrastructure; in such cases, the *lfn* denotes a fully qualified *Uniform Resource Identifier* (*URI*) to that file. To ensure the tracking of a dataset's ownership, an identification of the owner is also stored along with each record. Moreover, there may be cases where a physical dataset is located in a hierarchical data file system served by an HTTP server. In that case, the root *Uniform Resource locator (URL)* of the dataset is stored. Finally, each pipeline can have multiple algorithms in it and an algorithm can be used in multiple pipelines, the intermediary entity *Pipeline_has_Algorithms* maintains this relationship (as shown in Figure 2).

In the Analysis Base schema, *Subject* is the entity that holds information about the patient data stored in the Analysis Base. A subject can be associated with multiple *ImageFiles,* which stores references to subjects' *brain scans* stored on the N4U infrastructure*,* belonging to a *DataSet*. There is also a provision to bulk load image files in the Analysis Base with anonymised *Subject* information. In the Analysis Base schema the dataset definition starts with a *DataSetCategory* entity that defines each *DataSet*; for example, ADNI is a *DataSetCategory* and ADNI1, ADNI2 and ADNI-GO are various *DataSets* in the ADNI dataset category. Similarly, there can be multiple image files associated to each datasets. In addition to image files, the Analysis Base also stores a large number of clinical variables in the *AssessmentData* that is linked to an appropriate assessment type managed by the *AssessmentType* entity. At the time of writing this paper, over two hundred thousand image files and over 10 million clinical variables data have been indexed in this Analysis Base schema.

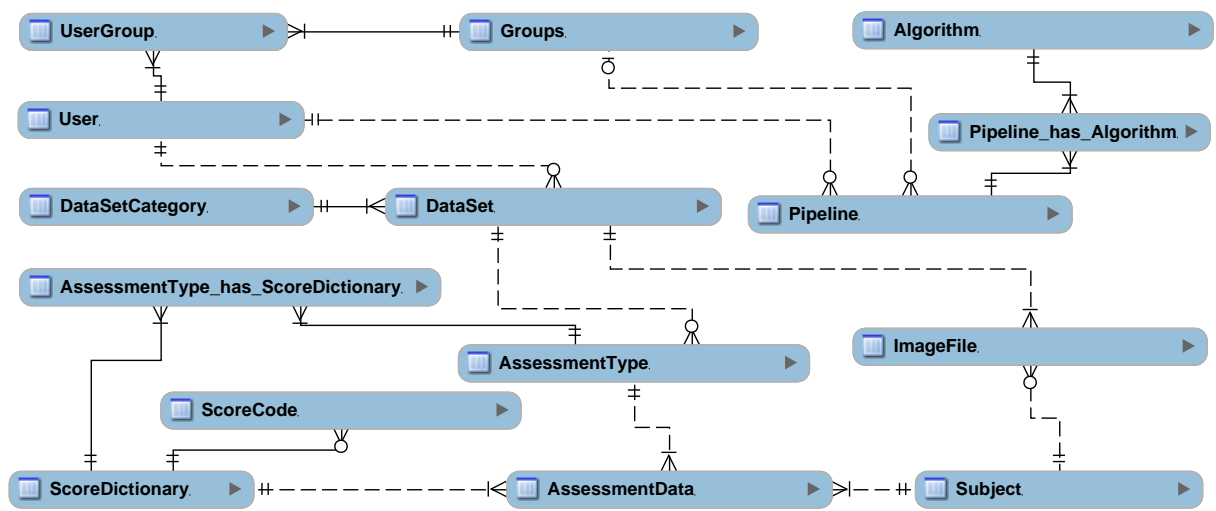

Figure 2: The Analysis Base schema diagram.

During brainstorming workshops of requirements specifications and refinements, we came across various end-users' information retrieval scenarios, where it was not obvious to the end-users how to provide values for the required clinical variables' query criteria. This means that they would not know the meaning of a variable or the type of values a specific variable has in the database. For example, in one dataset there is a clinical variable called *maritalstatus*, for which the database contains a number entries (also called scores) of 0, 1, 2, 3, 4, 5 and 9. Here, it was nearly impossible for the end-user to guess the appropriate matching number where "*Martial Status = Married*". In order to resolve the aforementioned scenario, a Clinical Variables' data dictionary has also been stored in the Analysis Base in the *ScoreDictionary* entity, which also contains clinical variables' description and/or associated questions. All possible values of clinical variables (available in the *ScoreDictionary*) are stored and linked (as an *one-to-many* database relationship) in the *ScoreCode* entity. More details on the use of data dictionaries are discussed in the Section 6 of this paper.

In order to populate the Analysis Base schema and to make this data available to the users of N4U services, a mechanism is required that is usable by the software components which will eventually also be used by human users. In the following sections, the N4U information services are presented that carry out all the necessary functions of *importing*, *indexing, storing and querying* of the datasets in and from the Analysis Base.

## 4. N4U Information Services

A set of integrated N4U Information Services has been defined as a key vehicle in meeting the intended objectives of the N4U project to provide access to the underlying N4U infrastructure and Analysis. The Information Services first catalogue the datasets stored on the Grid into the Analysis Base and provide access to clinical researchers through these indexes. Figure 3 illustrates the detailed role of the N4U Information Services in the N4U Virtual Laboratory setup.

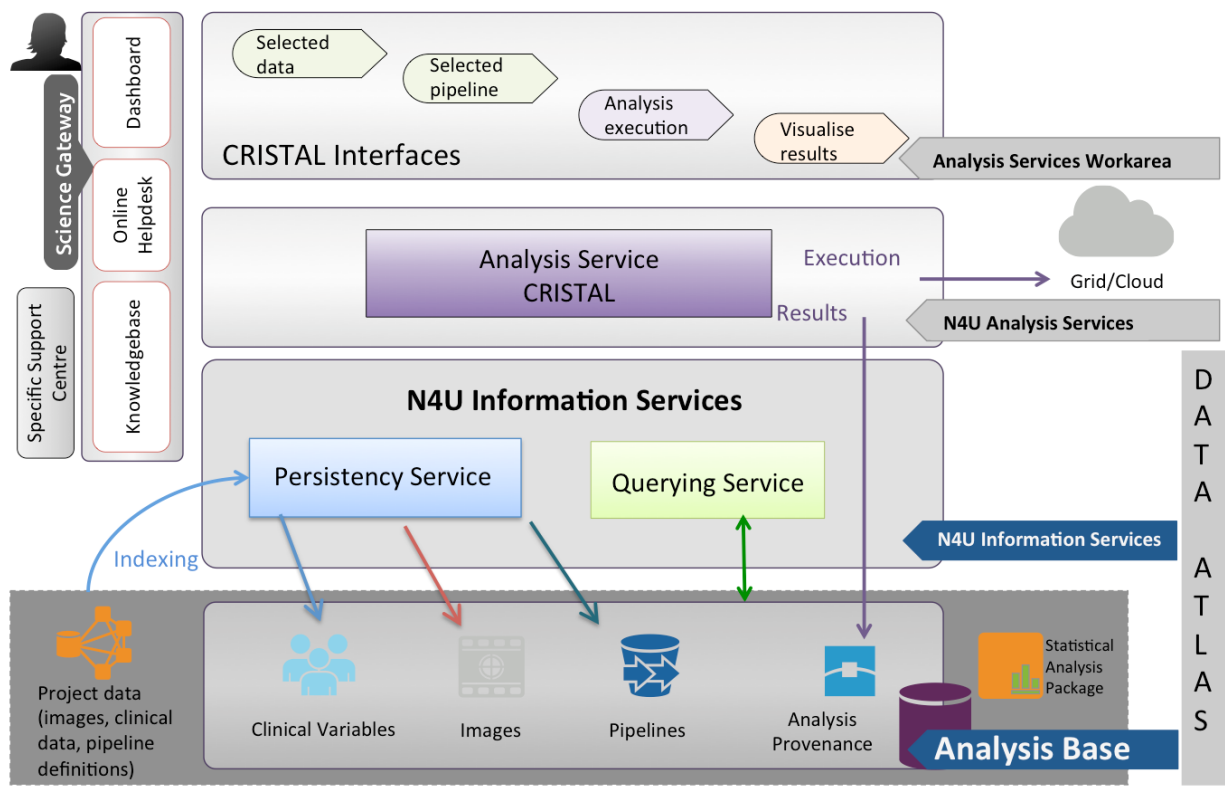

Figure 3: The N4U Virtual Laboratory architecture with essential components and services.

## 4.1. The N4U Analysis Base and Data Indexing Requirements and Challenges

The core data in the N4U infrastructure includes the clinical datasets from various neuroimaging studies and image files from scans carried out on patients as part of clinical studies. All datasets are imported from external sources into the N4U Grid-based storage and compute infrastructure. As noted earlier, filtering gigabytes of data at runtime is a non-trivial task, unless it is methodically indexed beforehand in the Analysis Base. The Persistency Service carries out all the necessary functions of *importing*, *indexing* and *storage* of the datasets in the Analysis Base. The indexing of datasets is carried out to meet the following identified requirements and associated challenges:

− Neuroscientists will use the N4U virtual laboratory to carry out analyses by executing pre-defined scientific pipelines (also known as workflows) over different datasets.

− The N4U Virtual Laboratory needs to provide a list of the pipelines that are executable on the infrastructure. The pipelines may have constraints such as being restricted to certain datasets (or formats) as inputs, algorithms that can be employed within those pipelines, etc. These constraints and other characteristics are usually specified in pipeline definitions. Therefore, the Analysis Base needs to index the pipelines registered in the N4U infrastructure. For completeness, other entities such as algorithms related to such pipelines also need to be indexed in the Analysis Base.

− In order to index the datasets stored in the N4U infrastructure, the datasets indexes need to be exported into the Analysis Base. Exporting the whole datasets into the Analysis Base is not feasible because the datasets are prohibitively large in size. For the purposes of the N4U Information Services, it is necessary to create indexes of the datasets in the Analysis Base. However, creating these indexes becomes challenging when the metadata associated with the datasets is sparse or unstructured.

− The above challenges are further compounded by the presence of the disparate formats of the various datasets used in the N4U project. If such metadata existed, preferably in a standardised format that could accommodate the differences in dataset formats, it could be exported into the Analysis Base for the creation of the indices in question. However, there is no standard metadata format across the range of datasets considered in the N4U project. Standardised interfaces for this purpose could also have been useful for accommodating any future datasets in the N4U infrastructure and their indexing in the Analysis Base.

The above-mentioned challenge of indexing heterogeneous datasets has been met by designing a generic data model for the Analysis Base (as discussed in Section 3). The Persistency Service indexes the data from the N4U's Grid-based storage and compute infrastructure into the Analysis Base using this data model. The functional perspective and other important details of the Persistency Service are explained in next sections.

## 4.2. The Indexing of Pipelines and Datasets via Persistency Service

The Persistency Service provides the following functionality:

− *Crawl* the datasets in the N4U Grid storage infrastructure; or parse the user dataset sent to the Persistency Service

− *Create* a software representation of the file system based storage structure that conforms to the data model of the Analysis Base;

- *Store* the clinical data (studies, results, clinical variables' values/scores etc.) associated with a dataset in the Analysis Base, such that these are query-able;
- *Index* the image files within a dataset in the Analysis Base such that they are associated with the clinical data of that dataset;
- *Index* the pipeline definitions (and algorithms) available in the N4U infrastructure in the relevant Analysis Base storage structures;
- *Store Metadata* associated with pipeline definitions identifying applicability of pipelines/algorithms to specific datasets.

The Persistency Service has been designed as a *web service* named PersistencyService deployed on the development gateways of the N4U Grid infrastructure. The PersistencyService exposes a set of operations that can be invoked by an administrator[1] or authorised N4U services[2] to execute its various functional tasks. Some of the main functional tasks of the PersistencyService are described as follows:

- **Pipelines/Algorithms Indexing**: The Persistency Service can be used to index the algorithm/pipeline definitions available in the N4U infrastructure.
- **Dataset Indexing**: To index a new dataset available in the N4U Grid storage, the administrator provides the Persistency Service with the directory structure of the dataset. Crawling through the directory structure of datasets can generate the required directory structure representation used by the Persistency Service. The service iterates through the dataset's constituent directories and builds a tree-like structure of all subdirectories and their associated contents.
- **Image Files Indexing**: In N4U, the brain images are stored in the DICOM format. The Persistency Service indexes all image files encountered during the iteration of a dataset's directories. The index is made up of a fully qualified lfn that points to the location of the image file in the Grid storage.
- **Clinical Variables Data Storage**: In N4U, the clinical and subject data (e.g., patient demographics) associated to neuroimages are provided in the comma separated values (CSV) format. The CSV format was used to meet the requirements of data provider who already had such clinical and subject data available in the CSV files. The data providers upload these files on the N4U Grid-based storage stored in the dataset directories. The Persistency Service parses the contents of these CSV files and stores them in the Analysis Base along with the lfn of images.

The following sub-section describes the process of indexing a dataset from the N4U Grid infrastructure in the Analyse Base with the help of a practical use case. As of writing, a number of datasets were available in the N4U Grid infrastructure that are indexed in the Analysis Base including the

---

[1] The PersistencyService is exposed as an interactive service to the normal users of the N4U virtual laboratory.

[2] For example, neuroscientist may create a sub-dataset to carry out an analysis via the N4U Analysis Services. Storing this user-defined sub-dataset is useful as it may speed up future analyses for the user or utilized by another user for verification of the neuroscientist's analysis. This storage may be carried out by one of the Analysis Services by invoking the relevant interfaces of the Persistency Service.

Open Access Series of Imaging Studies (OASIS www.oasis-brains.org) [25], with two categories i.e., Cross-sectional and Longitudinal, the Minimal Interval Resonance Imaging in Alzheimer's Disease (MIRIAD http://www.ucl.ac.uk/drc/research/miriad-scan-database), the Functional Bioinformatics Research Network (FBIRN fbirnbdr.nbirn.net:8080/BDR) Phase I and Phase II [26], the European Diffusion tensor imaging study in Dementia (EDSD) [27], the Magnetic Resonance in Multiple Sclerosis (MAGNIMS http://www.magnims.eu/), the Northwestern University Schizophrenia Data and Software Tool (NUSDAST http://niacal.northwestern.edu/projects/9) [28], the Alzheimer's Disease Neuroimaging Initiative (ADNI http://adni.loni.usc.edu/) datasets i.e., ADNI 1, ADNI 2 and ADNI GO, 1000 Functional Connectomes Project (1000FCP http://fcon_1000.projects.nitrc.org/), the Alzheimer's Repository Without Borders (ARWIBO http://www.arwibo.it/), the Autism Brain Imaging Data Exchange (ABIDE http://fcon_1000.projects.nitrc.org/indi/abide/), the Center for Biomedical Research Excellence (COBRE http://fcon_1000.projects.nitrc.org/indi/retro/cobre.html), the Attention Deficit Hyperactivity Disorder (ADHD-200 http://fcon_1000.projects.nitrc.org/indi/adhd200/) and the International Neuroimaging Data-sharing Initiative for Diffusion-weighted MRI (INDI_DWI http://fcon_1000.projects.nitrc.org/indi/indi_ack.html). For the purpose of describing the functional details of the Persistency Service, we have selected the NUSDAST and FBIRN dataset as demonstrative use cases in this paper. These selected datasets also demonstrate the structural differences in the data available in the N4U infrastructure.

## 5. Persistency Service Dataset Indexing Use Cases

This section presents use cases of the indexing of different types of datasets in N4U. These datasets vary in their directory structure on the Grid, and also in their internal structure – i.e., how they represent the medical data along with data dictionaries and image files. It is important to understand the file and directory structure of these datasets in N4U due to two main reasons. The first reason is to enable user access to the files stored on the N4U Grid infrastructure using the search mechanism provided in the Analysis Base. In order to achieve this, the Analysis Base should know about the location and path (i.e., lfn) of these files. The second contributing reason is related to the default representation of the dataset's metadata in N4U, which is in the CSV format. In many available clinical datasets in N4U, the CSV files representing the clinical variables and subject data of datasets do not maintain the image file lfns. However, they do keep track of the associated subject id. These scenarios presented a number of challenges for the Persistency Service's design and implementation and required devising a mechanism to link the clinical data from the CSV with the image files stored on the N4U Grid infrastructure.

## 5.1. The NUSDAST Dataset

The schizophrenia community has invested substantial resource on collecting, managing and sharing large neuroimaging datasets. Its efforts have resulted in a resource known as the Northwestern University Schizophrenia Data and Software Tool (NUSDAST) that provides high-resolution magnetic resonance (MR) data from subjects with schizophrenia. The subject data also provides information about the non-psychotic siblings and their health control parameters. The NUSDAST dataset has the following two notable properties: (i) it combines the neuroimaging data with the demographic, clinical, neurocognitive and genotype information; and (ii) it consists of 368 subjects and for each subject, there is an arbitrary number of files (minimum 3 to maximum 33), with different characteristics such as MPR, FLSH etc.

All datasets to be utilised in the N4U project are stored in the *data* directory of the N4U Grid infrastructure. These datasets contain both clinical study data and images. All clinical study related data (called clinical variables) are made available as variable length comma separated values (CSV) in the CLINICAL_VARIABLES directory. The IMAGES folder contains the image scans taken as part of the clinical studies in various formats (e.g. NIFTI .nii format). These images are organised in sub-directories that are named after the subject IDs and can thus be cross-referenced with the clinical study data of the subjects. Figure 4 illustrates the directory structure of the NUSDAST dataset in the N4U Grid storage. The .csv files in CLINICAL_VARIABLES directory provides information about the data dictionary, which provides metadata information about the clinical variables, and actual clinical data, which includes each subject's clinical, neuropsychological and socio-demographical variables.

Each images subdirectory is named after the scanned patient's id. The structure of scanned image files inside the subject/patient sub-directories e.g. nG+NUSDAST+CC0196, is as follows:

```
IMAGES/nG+NUSDAST+CC0196/
├── nG+NUSDAST+CC0196+M0+1T5+3DSF+ORIG+V01.tar.bz2
├── nG+NUSDAST+CC0196+M0+1T5+FLSH+ORIG+V01.ifh
├── nG+NUSDAST+CC0196+M0+1T5+FLSH+ORIG+V01.nii.bz2
├── nG+NUSDAST+CC0196+M0+1T5+MPR1+ORIG+V01.nii.bz2
├── nG+NUSDAST+CC0196+M0+1T5+MPR2+ORIG+V01.nii.bz2
├── nG+NUSDAST+CC0196+M0+1T5+MPR3+ORIG+V01.nii.bz2
├── nG+NUSDAST+CC0196+M0+1T5+MPR4+ORIG+V01.nii.bz2
├── nG+NUSDAST+CC0196+M0+1T5+MPRA+PROC+V01.nii.bz2
├── nG+NUSDAST+CC0196+M0+1T5+MPRA+PROC+V01.rec
…
…
```

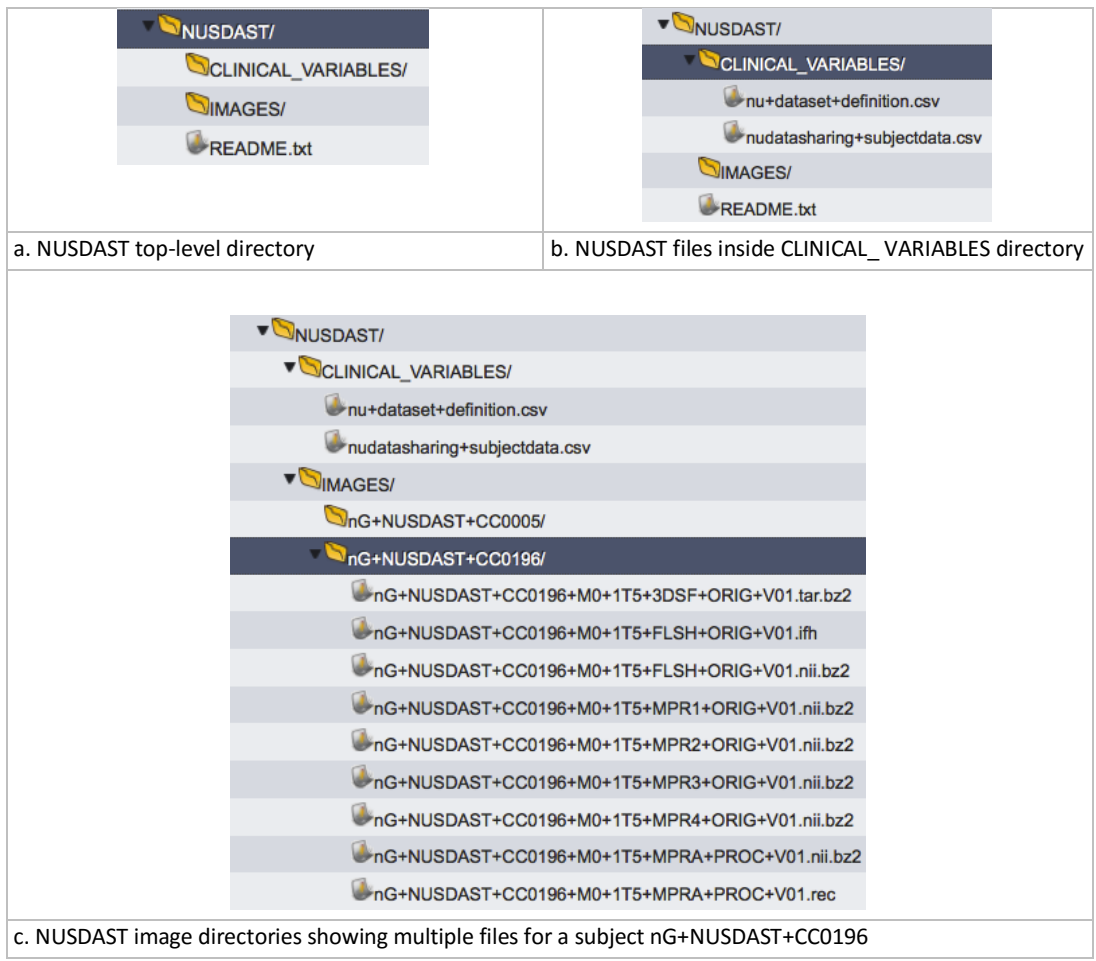

Figure 4: NUSDAST dataset directory structure in the N4U Grid storage infrastructure.

Note that in the patient's id directory (e.g. nG+NUSDAST+CC0196 in Figure 4) the files with extensions *.rec* and *.ifh* represent the summary information of all the scans. The first 4 scans are MPRAGE acquired at 1.5T, while the last one is the average over the first 4 scans to improve the signal-to-noise ratio (SNR). The naming convention adopted for the NUSDAST scan image files is explained as follows:

- **nG** = neuGRID
- **NUSDAST** = NUSDAST dataset
- **CC0196** = patient's id
- **M0** = Month 0 (up to 3 time points: M0, M24, M48)
- **1T5** = Field strength
- **3DSF** = 3D surface object or FLSH = FLASH or MPR1/2/3/4 = MPRAGE acquisitions or MPRA = MPRAGE average
- **ORIG/PROC** = original or processed scan
- **V01** = it indicates the version of the scan.

## 5.2. The FBIRN Dataset

The Biomedical Informatics Research Network [29], FBIRN dataset is hierarchical in nature, representing different phases such as Phase I and Phase II. The Phase I Traveling Subjects study was the first fBIRN multi-centre study. Five healthy subjects were imaged on two occasions on 9 or 10 different scanners located in geographically diverse locations. The purpose of this study was to provide a reference dataset with which to assess test-retest and between-site reliability of *fMRI* - Functional Bioinformatics Research Network [14] - and to provide a rich dataset to test tools and methods to allow for the calibration of between site differences in the fMRI results.

Each phase in FBIRN has its own sub-assessments as shown in the directory structure in Figure 5 (b). The .csv files representing clinical variables and metadata information are placed inside sub-directories as shown in Figure 5(c).

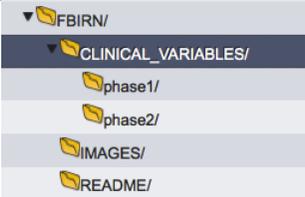

| | |
|---|---|
| a. CLINICAL_ VARIABLES structure of FBIRN dataset | b. Each phase has its own set of assessments inside CLINICAL_ VARIABLES directory |
| c. FBIRN clinical variable CSVs for ASI assessment | d. FBIRN image directories of subjects in Phase I for each center |

Figure 5: FBIRN dataset directory structure in the N4U Grid storage infrastructure.

Images in FBIRN are also arranged in an hierarchical manner representing different phases such as Phase1 and Phase2. The phase directory has further sub-directories representing the centres where the scans were performed. Different scan images of a subject are placed inside a subdirectory named after the scanned patient's identification (id). The structure of scanned image files inside the subject/patient sub-directories e.g. nG+000900000106, is as follows:

**IMAGES**/phase1/CENTRE0003/nG+000900000106
├── nG+FBIRN1+000900000106+1T5+BH1+ORIG+V01.tar.bz2
├── nG+FBIRN1+000900000106+1T5+BH1+ORIG+V02.tar.bz2
├── nG+FBIRN1+000900000106+1T5+BH2+ORIG+V01.tar.bz2

```
├── nG+FBIRN1+000900000106+1T5+BH2+ORIG+V02.tar.bz2
├── nG+FBIRN1+000900000106+1T5+MMN1+ORIG+V02.tar.bz2
├── nG+FBIRN1+000900000106+1T5+MMN2+ORIG+V01.tar.bz2
├── nG+FBIRN1+000900000106+1T5+MMN2+ORIG+V02.tar.bz2
├── nG+FBIRN1+000900000106+1T5+MN1+ORIG+V01.tar.bz2
├── nG+FBIRN1+000900000106+1T5+MPR+ORIG+V01.tar.bz2
├── nG+FBIRN1+000900000106+1T5+R1+ORIG+V01.tar.bz2
├── nG+FBIRN1+000900000106+1T5+R1+ORIG+V02.tar.bz2
├── nG+FBIRN1+000900000106+1T5+R2+ORIG+V01.tar.bz2
…
…
```

Note that in the patient's id directory (e.g.: nG+000900000106) there are multiple files. The naming convention adopted for the FBIRN phase1 scan image files is explained as follows:

- **nG** = neuGRID
- **FBIRN1** = Dataset name
- **000900000106** = patient's id
- **1T5** = Field strength of 1.5 Tesla or 4T which means 4.0 Tesla
- **SM2** = It is the acquisition modality. Other modalities such as BH1, BH2, MMN1, MMN2, MPR, R1, R2, SIRP, SM1, SM2, SM3, SM4 and T2 are also possible.
- **ORIG** = original or processed scan
- **V01** = it indicates the version of the scan.

It is apparent from the above discussion that there are different types of datasets, which vary in structural representation on the Grid and clinical data. The FBIRN dataset is arranged in three sub-levels, the OASIS dataset is arranged in two sub-levels and the NUSDAST dataset is arranged in one sub-level only. Since there is no uniform structure across all the datasets, it is a challenge to devise a flexible mechanism that can accommodate such diversity in current datasets as well as future datasets. Based on this information and understanding, the Persistency Service devises a mechanism backed by the supporting (Analysis Base's) storage design which can address this challenge. Given the above on-disk structure of the datasets in the Grid storage, their indexing in the Analysis Base by the Persistency Service is described in the next section.

## 5.3. Indexing the NUSDAST Dataset in Analysis Base

From a software design point of view, the PersistencyService consists of a crawler, parser and model components that represent the datasets according to the database schema. The crawler browses through the directory structure of a dataset and records the contents in a tree-like structure. The parser component parses the directory structure along with data dictionary files associated with a dataset. The model component transforms the clinical variable data into insert-able SQL (Structured Query Language) objects by linking clinical variables with image lfns and patient ids. The SQL objects are mapped directly to the relevant tables of the Analysis Base. Figure 6 illustrates the Persistency Service's functional flow during the indexing of a dataset from the Grid storage into the Analysis Base. The details of this process are as follows.

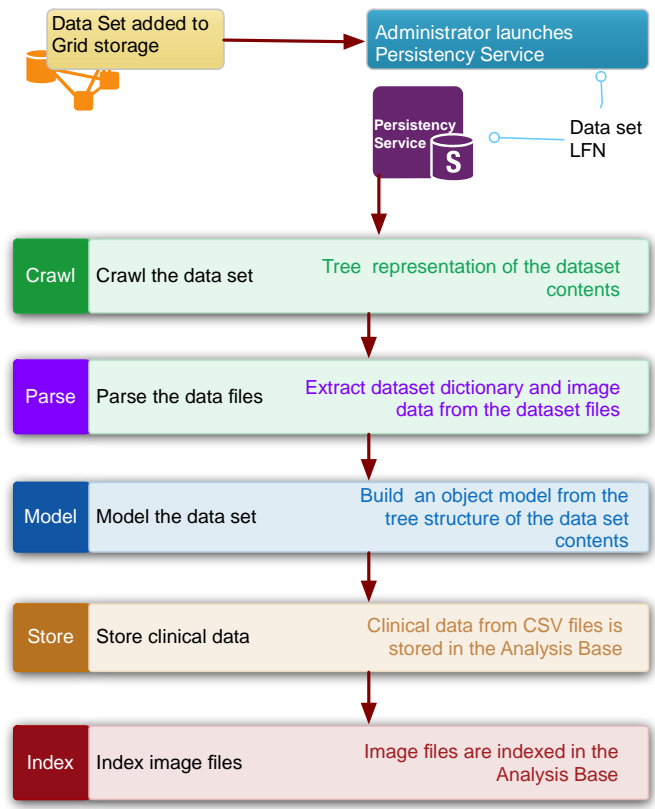

Figure 6: The Persistency Service's functional flow during the indexing of a dataset from the Grid storage.

The crawler creates the Grid directory structure of a given dataset. This directory structure represents the image directory and image names in each directory. This information is passed on to the parser. The parser analyses the directory structure and extracts the image lfns and the associated patient ids. It also parses the dataset dictionary, which describes the metadata of the clinical variables and possible score values (if any) for each clinical variable in the given dataset. The metadata information that is extracted from the dictionary files is stored in the Analysis Base. The image lfns, patient id (subject id) and clinical study data (in a CSV format) are passed to the model component (step 3 in Figure 6). The model iterates over the clinical study data file and retrieves a list of clinical variable values, collects subject information, and creates a mapping between a subject and its image files (passed from step 2). Once all this information is available, the PersistencyService then creates SQL models for each record according to the relevant tables that can be inserted in the Analysis Base. Figure 7 shows the Persistency Service's flow chart diagram illustrating the flow of activities during data set indexing in the Analysis Base.

The indexing of image files in the Analysis Base requires appropriate relationships to be established between each image file (its lfn to be precise) and the clinical variable's values. As explained in Sections 5.1 and 5.2, the names of the sub-directories containing the individual image files and the image file names contain the subject IDs (i.e., the anonymised patient names). This naming convention aids in appropriately indexing the image files in the relational database model of the Analysis Base. Once the PersistencyService indexes a dataset (in terms of subjects, dataset dictionary, clinical variables and images) in the Analysis Base, other N4U services, such as the Querying Service, are able to utilise these indexes in order to access the Grid-stored datasets instead of searching the full Grid-storage at runtime.

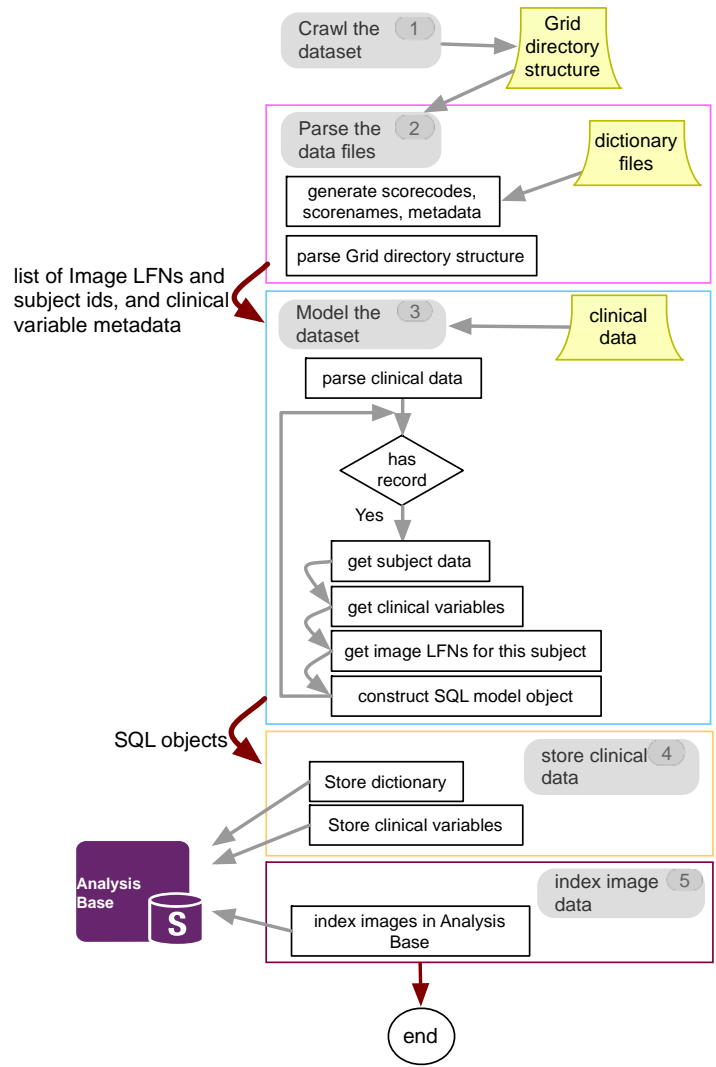

Figure 7: The Persistency Service's flow chart diagram illustrating the flow of activities during data set indexing.

Before discussing how other services can access the datasets via these indexes (Section 6), the following subsection briefly describes the indexing of algorithms and pipelines in the Analysis Base by the PersistencyService.

## 5.4. Persistency Service Pipeline of Algorithms Indexing Use Case

In addition to indexing datasets stored in the N4U Grid infrastructure, another function of the Persistency Service is to index pipelines of algorithms, in the form of scripts, which are also stored in the Grid infrastructure. The Persistency Service indexes the available pipelines of algorithms so that the users can make appropriate selections when defining their analyses. The following paragraph describes the mechanism used to index pipeline with its associated information.

In order to provide an exploratory and customizable search mechanism (discussed in Section 6), the Persistency Service indexes the pipeline and its associated information. Algorithms are also conceptually contained in individual pipelines i.e., the Persistency Service indexes the algorithms while maintaining the relationship between algorithms and their encompassing pipelines. Figure 8 shows the main steps in indexing a pipeline in the Analysis Base.

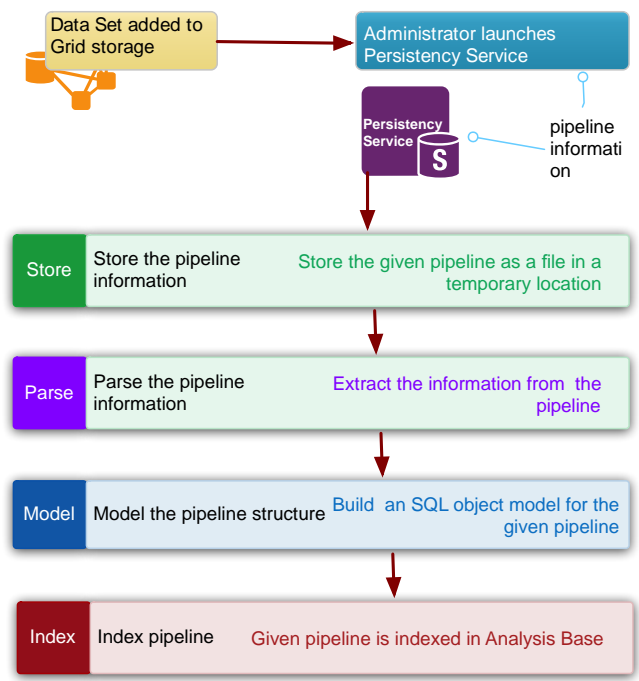

Figure 8: Main activities of the Persistency Service involved in indexing a pipeline.

After receiving the pipeline information, it is stored in a temporary file on the server. The pipeline information is then parsed to extract its name, lfn, version, description, algorithm information, etc. The parsing is performed in order to avoid null or empty fields and to build SQL objects with appropriate values. The SQL objects transform the given pipeline information onto the underlined schema in order to store the pipeline and its associated information. Once the SQL objects are created, the pipeline information is stored and indexed in the Analysis Base and then exposed to the Explorer interface, which is a part of Querying Service (discussed in Section 6.2).

## 6. Analysis Base Information Retrieval through the Information Services

In order for N4U users – and other N4U services – to be able to retrieve information from the Analysis Base, a mechanism was required to search and query the data stored in it. This retrieval mechanism enabled users to perform their queries on clinical data and images, and to retrieve desired data related to pipelines of algorithms. Various web interfaces, called the Analysis Base Querying Interfaces, have been deployed to fulfil these requirements.

The Querying Service Interfaces enables N4U users to submit queries for information retrieval from the Analysis Base; for example, the N4U datasets' (listed in Section 5) related information and/or the meta-data associated with an image file stored in the Analysis Base. There is a wide range of queries that is supported through the querying interfaces, including browsing the datasets, pipelines and algorithms, viewing data dictionary, searching for the metadata in the patients' clinical data etc. These queries may arise from different such as from the set of N4U services (as shown in Figure 9) and thus the Querying Service exposes different interfaces for different clients. A summary of different functions offered by the various Interfaces of Querying Service is listed below:

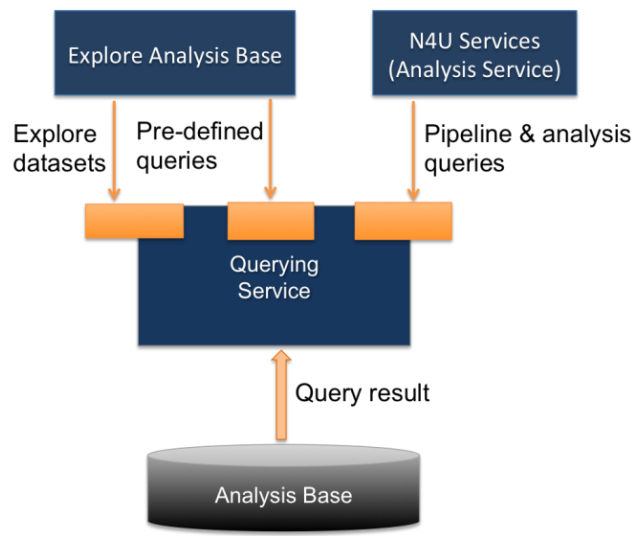

Figure 9: Users and other services accessing information using the Querying Service interfaces.

- Using the Querying Service, users can search within multiple datasets based on different search criteria and use the resultant data in their analyses.
- A few examples of supported datasets queries include: (i) searching for all or specific datasets; (ii) getting a dataset identified by an unique id; (iii) searching in a dataset based on the given values of various clinical variables; and (iv) retrieving the lfn location of an image, etc.
- A user can search for and catalogue a particular subset of image and data files residing in the N4U Grid infrastructure. This function allows for an easy-to-use mechanism for the users to access the information in the Analysis Base from a single point of access.

- Performing exact match and SQL *'like'* operations for the filtering of clinical variable data. Users can also perform comparison operations (i.e., >, <, = etc.) on clinical variables' values in the visual Query Builder.
- A user can navigate into a dataset (or subsets) and view all the clinical variables linked to that dataset. A user can also view the metadata and/or detailed data dictionary associated with each clinical variable.
- The ability to export the output of user queries/inquiries about clinical data and images as both xml and csv files, which allows subsequent processing of data in other applications (e.g. excel and analysis service) and also import into user-owned databases.

In the following subsections we discuss these points in more detail along with the design and implementation aspects of the Querying Service's sub-components.

## 6.1. Querying Service Design and Implementation

The Querying Service is designed as a web service to achieve its multiple objectives. A web service, invoked remotely over HTTPS, allows us to expose the desired functionality in a controlled manner for the client applications. All the functionality embedded at the server-side not only provides a transparent access to the service functionality but also requires a small footprint at the client side. This setup allows for a rich and easy to use but controlled functionality for the users and external services. The N4U user or other services interact with the Analysis Base through a service-oriented approach that is illustrated in Figure 10.

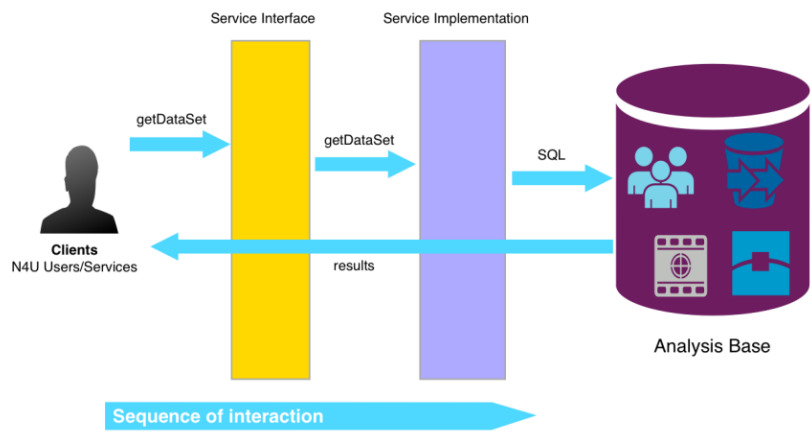

Figure 10: The user interaction with the Analysis Base through a service-oriented approach.

The Querying Service has been developed using the Apache CXF framework (http://cxf.apache.org/). The Querying Service is only accessible over HTTPS on the N4U gateway and only valid N4U users with active, authenticated sessions are allowed to access this service. Its implementation makes use of a user identity retrieval mechanism provided by the N4U gateway's runtime environment

(https://neugrid4you.eu/web/science-gateway), which authenticates all incoming requests. This approach provides two advantages; (i) no non-legitimate request can access the N4U data, thus making it secure and (ii) the user identity is used to log and retrieve information stored by that user in the Analysis Base. This request-response flow is illustrated in Figure 11.

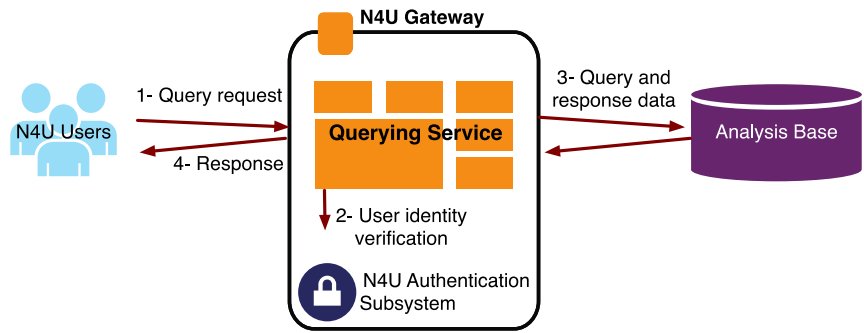

Figure 11: Flow of activities in the Querying Service.

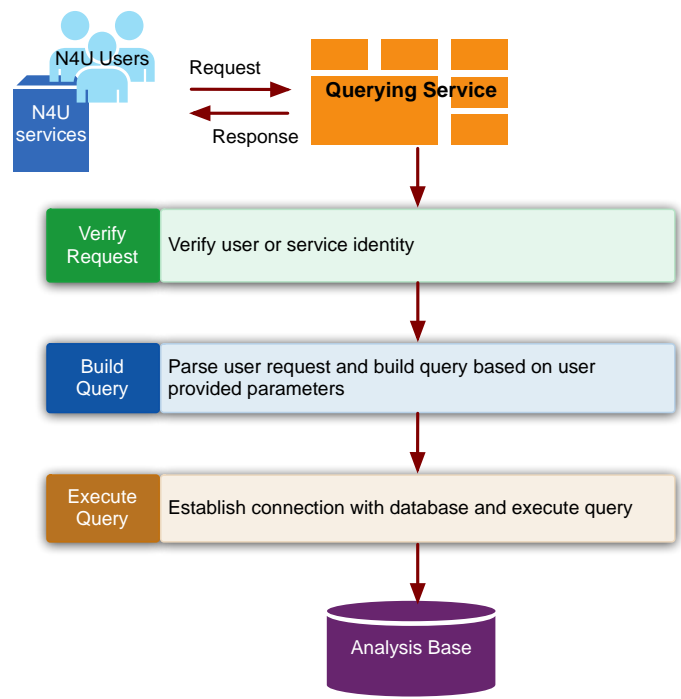

Figure 12: The Querying Service's functional flow in handling a query request.

After the successful verification of each user request, the Querying Service starts building the appropriate search query against the user parameters provided in the request. In the query-building phase, user selected parameters are parsed, and the required tables and appropriate joins are selected based on the given parameters. The Querying Service then connects to the backend database and executes the query. The query result is retrieved from the database and presented to the user. In the case of failures e.g. user verification or query errors, human readable and understandable messages are sent back to the user. The internal flow of activities within the Querying Service is shown in Figure 12.

In order to test the Data Atlas functionality using the Querying Service interfaces for various types of users, a dedicated work package has been designed in N4U to specify user requirements and evaluation framework – using User Acceptance Testing (UAT) - that describe user requirements and also verify the developed services. The UAT has been performed in 43 sessions in different locations such as Stockholm, Brescia, Amsterdam, Kuopio, Munich, Warsaw and Bern in Europe with the help of different types of N4U user group such as Neuroscientist, Pharma, Developer and Administrator. These users are divided into internal and external to the N4U infrastructure categories. The internal users are the ones who are participants or developers of the N4U services, and the external users are mainly the neuroscience and pharma researchers. The details about user requirements, evaluation framework, trainings and user acceptance testing are presented in the relevant project deliverables [30]. The results of the UAT confirmed that all of the essential requirements such as interface for basic search and interface for advance search related to the Querying Service and Data Atlas have been successfully achieved. The following sections discuss different analysis base querying interfaces provided by the Querying Service.

## 6.2. Information Retrieval via Analysis Base Explorer

While designing various Analysis Base Querying interfaces, special focus was placed on increasing the ease of learning for novice users as well as providing intelligent dataset filtering features for expert and advanced users. The main purpose of having a querying mechanism on top of the Analysis Base was to facilitate a user or external services in fetching the data stored in it. However, not every user – particularly a neuroscientist - is equipped with database querying skills and relevant technical expertise. Because of this, a feature rich web interface has been designed and developed to present the N4U Analysis Base to the N4U community. The *Analysis Base Explorer* presents an accessible and easily usable view of the N4U Analysis Base from the perspective of a naive user. It is designed to showcase the main datasets, sub-datasets and pipelines stored in the Analysis Base. When a user accesses the Explorer page, two main entities i.e., the datasets and pipelines are presented. Upon selecting one of these entities, an AJAX (Asynchronous Java and XML) request is sent to the Querying Service to retrieve further information (data) about the selected entity. The resultant information is displayed which also includes the total number of records as well as the query response time, and this further improves the users' experience. Through this Explorer interface, users can also observe the relationship between pipelines and algorithms. For instance, when a user selects Algorithms after selecting a Pipeline, the data for the relationship between the Pipeline and the Algorithms are also displayed. This kind of information is particularly useful for neuroscientists looking for a specific algorithm in a particular pipeline.

Since data stored in the Analysis Base can massively grow over time and the links between different entities can also cause a large amount of data to be retrieved, this presented a challenge in retrieving and presenting the data at the client side. To safeguard against this potential problem, a pagination technique has been adopted through which the data is divided into a configured number of records (set to 300 records per page by default) when presented to the user. This not only decreases the burden on the backend database server answering user queries but also significantly reduces the response time and rendering time at the client/user side.

### 6.3.    Information Retrieval via Analysis Base Querying Builder

In the previous section, we have discussed how the *Explorer* component of the Querying Service assists novice users in viewing the data stored in the N4U Analysis Base. A user, such as a neuroscientist, may also want to retrieve the neuroimaging data, filtered on the basis of certain clinical variables. To support this type of complex search and selection, a dynamic and user driven *Query Builder* has been incorporated into the web interface. The Query Builder allows a user to graphically design queries based on her selection criteria to search for the images within the datasets indexed in the Analysis Base. In order to further facilitate N4U users, three different types of graphical query building methods are provided: (a) Predefined queries, (b) Clinical variable search; and (c) Advanced search (as shown in Figure 13). These methods are explained in following subsections.

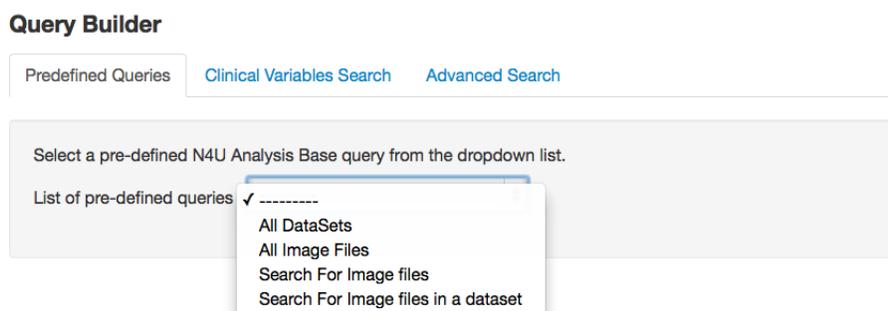

Figure 13: Query Builder interface to design queries and retrieve data from the Analysis Base.

### 6.3.1.   Predefined Queries

In the Predefined Queries tab, a pre-populated list of predefined queries is provided to the user to fulfil basic users' queries. The reason for having such a list of predefined queries is to accommodate a number of common use cases and to save users' time in creating their own queries for common scenarios. A dropdown list of queries is given to the users from which they can select an appropriate query. A brief description of these queries is given below.

- *All DataSets:* This query fetches a list of datasets and their associated metadata. The metadata contains dataset name, id, *lfn*, owner, creation date etc.
- *All ImageFiles:* This query retrieves a list of entire image files indexed in the N4U Analysis Base. Since there could be a large number of image files, the result is divided into pages for performance reasons as discussed earlier in Section 5.2.
- *Search for Image files:* In this option, a text field is presented to the user in which she can provide an image filename or part of an image's *lfn*. This query searches for the given input in image names or *lfns* in all datasets and retrieves a list of matched image files.
- *Search for Image files in a dataset:* In this option, the user provides the image name or its *lfn* along with dataset type in which she wants to look for the given image name or *lfn*. The Querying Service looks for the given image file in the provided dataset only.

The list of predefined queries is easily extendable and administrators can add further queries for common scenarios.

### 6.3.2. Clinical Variable based Search

For the clinical variable based search, the Query Builder takes user supplied parameters in a stepwise manner. Firstly, it shows a list of available datasets to the user. Each of the datasets can have multiple subcategories or sub-datasets. Upon selecting one of the datasets, the Query Builder dynamically loads the associated sub-datasets e.g. the OASIS-CrossSection and OASIS-Longitudinal for the OASIS dataset. Each subtype can have a different number of clinical variables, possibly with different names. Upon subtype selection, its associated clinical variables are dynamically created and presented as a dropdown list. In N4U, it is also possible that a dataset such as NUSDAST does not have further subcategories. For such a dataset, clinical variables associated with it are loaded and presented in a dropdown list to a user. Neuroscientists can then select from the available clinical variables and provide the selection criteria by filling the text fields. The interface allows a user to specify a multi-parameter query by selecting and providing search criteria for multiple clinical parameters. The user may also remove certain parameters while the query building is in process or restart from scratch. As shown in Figure 14, all of the associated clinical variables for the user selected dataset 'NUSDAST' are shown to the user in order to perform a selection of multiple clinical variables and to create an appropriate search filter.

Figure 14: Clinical variables-based parameterized querying and the use of the Data Dictionary.

### 6.3.3. Use of Metadata and Data Dictionary for Clinical Variables based Search

While implementing and testing the Clinical Variable search, we came across various scenarios where it was not obvious for the end users how to provide values for the selected clinical variables. This means that they would not subsequently know the meaning of the clinical variable or the type of values that the selected variable has in the database. For example in NUSDAST and ADNI datasets there is a clinical variable called *maritalstatus* i.e., "Martial Status", for which the database contains number entries as 0, 1, 2, 3, 4, 5 and 9. Here, it was nearly impossible for the end user to guess the appropriate matching number where "Martial Status = Married".

In order to resolve the aforementioned scenario, Clinical Variable data dictionaries have also been stored in the Analysis Base and their relationship has been established with each Clinical Variable for all datasets. Due to this feature, when a user clicks on a Clinical Variable such as "Martial Status", the respective data dictionary values are automatically shown to the user as follows (see Figure 14):

*Clinical Variable: maritalstatus*

*Desc./Question:*

*0 ='Other'*

*1 ='Single'*

*2 ='Married/common law'*

*3 ='Divorced'*

*4 ='Separated'*

*5 ='Widowed'*
*9 ='Unknown'*

With this, if the user is interested in only retrieving those subjects with *Marital Status='Married'*, then she can see from the data dictionary entries and enter 2 in the text field (as shown in Figure 14). Similarly, for another similar Clinical Variable "*employmentstatus*" (i.e., Employment Status), the following are the associated data dictionary entries:

*Clinical Variable: employmentstatus*
*Desc./Question:*
*0 ='Other'*
*1 ='Employed full-time'*
*2 ='Employed part-time'*
*3 ='Unemployed'*
*4 ='Homemaker full-time'*
*5 ='Student full-time'*
*6 ='Student part-time'*

Sometimes there is additional linked metadata information to a clinical variable e.g. *Variable Type, Score Values and/or Comments*. In order to view such detailed information associated with a clinical variable, user can click on the "?" button (as shown in Figure 14), which is provided on the right-hand side of each selected variable, to display the complete metadata stored in the Analysis Base.

### 6.3.4.  Advanced Search for Expert users

While building queries/filters using the Query Builder interface, a user can use various types of operators such as = (*equal to*), < (*less than*), > (*greater than*), *Like* (for sub-string matching), " " (for exact string match), etc. Here the *equal* operator '=' compares single values to one another in a SQL statement. The *equal* sign (=) symbolizes equality. When testing for equality, the compared values must match exactly or no data is returned. Similarly the < (*less-than*) and > (*greater-than*) are used with numerical values, the *Like* operator can be used to treat the given text input as a sub-string to be matched anywhere in the clinical variable value. Within the Analysis Base web application detailed instructions have been provided on how to use these operators; users can also consult the public training videos provided as part of the learning material [31] to use such functions.

In addition, a user can easily specify more complex criteria (e.g. use of *OR* and *NOT Equal To, NOT IN* operations etc.) by copying the previously generated SQL and using it in the Advance Search tab. This feature is for the advanced users who have an understanding of databases and know how to specify a SQL query. A text area is provided where they can specify database queries to fulfil their requirements. In order to facilitate users, a mechanism is devised which enables users not to write queries from scratch; rather they can copy their previously formulated query via *Query Builder* and modify it within the *Advance Search tab*. In this way, it reduces the chances of an error, and takes less time and effort on the user's part to edit and/or extend a search criterion. Moreover, a user can only change the *WHERE* clause to modify the given filter or introduce further filtering (using OR, NOT etc. operators) as well as including further clinical variables in the overall search criteria. To avoid SQL injection [32] and other write-able queries such as INSERT, UPDATE or DELETE, the Querying Service executes these queries within a sandbox which is created by parsing the given query and using the read-only permissions on the database.

### 6.3.5. Exporting query results in XML and CSV format

In the above sections, we have listed a subset of the various mechanisms for users to locate their desired datasets in the Analysis Base. Another important feature of Analysis Base querying interfaces that this section describes is the exporting of search results or resultant outcome in XML or CSV formats, which can then be used by neuroscientists in any preferred neuro-science data analysis application(s).

Using the aforementioned Query Builder interfaces, once the user has entered search criteria and located the desired datasets, in addition to displaying the querying result, the user is also provided with the options of exporting the outcome in both CSV and XML formats. Users can do this without navigating away from the Query Builder interface and can generate a number of exports by just changing the filtering criteria. In order to enable fast data retrieval and the generation of XML and CSV, secondary B+ trees and bitmap indexing have been implemented on the data stored in the Analysis Base. When a user submits a search query based on the specified criteria, the retrieved outcome is displayed and both of the CSV Export and XML Export buttons appear automatically at the bottom of the page. Upon clicking the export buttons, an export request is sent to the server. The server processes the request, generates and transforms the resultant data in the requested format and returns it as a file (.csv or .xml). Figure 15 shows the generated and exported XML file. Each generated file is time stamped, which is unique for all files. Therefore the old generated files are not automatically replaced with new files. The generated XML structure provides information about an image file and its metadata, dataset and assessment, and its associated clinical variables, which were provided in the query. The following example shows one XML record from the exported XML in which individual records are separated with the *<Record> and <\Record>* tags, and data for the clinical variables are enclosed within individual named tags.

```
<Record>
    <imagefile_name>nG+NUSDAST+CC7959+M0+1T5+3DSF+ORIG+V01.tar.bz2</imagefile_name>
        <lfn>/grid/vo.neugrid.eu/data/NUSDAST/IMAGES/nG+NUSDAST+CC7959/nG+NUSDAST+CC7959+M0+1T5+3DSF+ORI
G+V01.tar.bz2</lfn>
        <imagefile_type>.bz2</imagefile_type>
```

```
        <imagefile_description></imagefile_description>
        <added_on></added_on>
        <dataset_id>32</dataset_id>
        <subject_id>nG+NUSDAST+CC7959</subject_id>
        <assessment_type>NUSDAST</assessment_type>
        <maritalstatus>4</maritalstatus>
        <race>2</race>
        <gender>male</gender>
</Record>
```

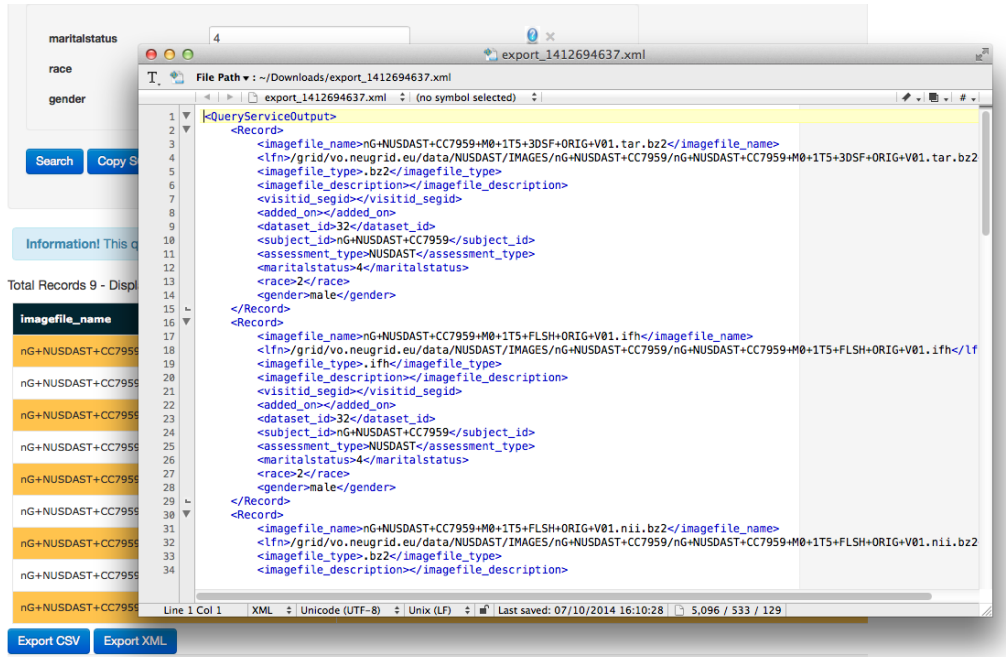

Figure 15: An example of XML export for the filtered datasets.

Whenever a user exports the query result in an XML format, seven (7) default values appear in all the records. These seven values are *imagefile_name, lfn, imagefile_type, imagefile_description, added_on, dataset_id, subject_id,* and *assessment_type*. Here *imagefile_name, lfn,* and *subject_id* are the most important variables in order to locate a particular image on the N4U Grid infrastructure. The rest of the clinical variables, which become part of the exported XML, depend on the user provided query criteria. As discussed above, a user can include unlimited numbers of clinical variables as search filters and accordingly all of them appear in the exported file. Finally, the outcome exported in the format of XML and CSV can be used by other applications to generate/perform an analysis or to retrieve an image or set of images from the Grid.

## 7. Conclusions

Recent developments in data management and imaging technologies have significantly affected diagnostic and extrapolative medical research and biomedical researchers are faced with severe problems of heterogeneous clinical data management. The impact of these new technologies is largely dependent on the speed and reliability with which the medical data can be visualised, analysed and interpreted. Grid computing, and recently Cloud computing, has alleviated some of the issues associated with the capture, processing and storage of huge numbers of medical images and associated clinical information. However, there is still a lack of clinician-friendly graphical interfaces and tools that provide an easy access to the (Grid) infrastructure-resident data. Moreover, there are no or few links between images, data dictionaries, clinical data and metadata associated with images, which makes it extremely difficult for biomedical researchers to define and conduct their analyses.

In order to address these issues, the N4U Virtual Laboratory presented in this paper focuses on providing an *intuitive, fast and linked* access to large-scale heterogeneous clinical datasets to derive a greater understanding of the neuro-degenerative diseases through data analysis. The N4U Data Atlas within the N4U Virtual Laboratory addressed the research and practical challenges by offering an integrated medical data analysis environment to optimally exploit neuroscience pipelines containing algorithms, large image datasets and clinical variables data in order to conduct analyses. The Analysis Base enabled such analyses by indexing and interlinking the neuroimaging and clinical study datasets, and pipeline definitions stored in the N4U Grid infrastructure. Furthermore, this paper has described the requirements, specification, implementation and deployment of the N4U Information Services i.e., (1) the indexing of pipelines/algorithms and heterogeneous datasets in the Analysis Base by the Persistency Service, with a particular emphasis on how the datasets that are stored in the N4U Grid-based storage infrastructure are indexed homogeneously in the Analysis Base; and (2) the retrieval of indexed information from the Analysis Base through the Querying Service; for which various (including dynamic) interfaces and methods are exposed to provide Analysis Base access to the end-users and other N4U services.

The main challenges faced and tackled in providing information services relate to the very large datasets sizes, different formats, structures and semantics of datasets. Moreover, the manner in which relationships between clinical parameters in CSV files and image files are maintained also varies across datasets. Moreover, unstructured data dictionaries compound the problem. To overcome these challenges, a generic schema model of the N4U Analysis Base has been presented; and a Persistency Service has been presented to index and link such data in the Analysis Base. While doing so, it has been specifically ensured that the users, and other N4U services, are provided with an uniform view of the datasets and data dictionaries in order to apply filters on available clinical parameters and retrieve relevant information.

In the future, this work can be extended to resolve various challenges associated with the management of large clinical datasets. As the volume and variety of datasets in terms of different formats and structures, can increase in future the domain becomes a *Big Data* candidate. Therefore, the use of *NoSQL* databases to provide fast and scalable storage and retrieval mechanisms along with biomedical Big Data analytics and information visualization can be explored. Moreover, the use of ontologies with domain knowledge on top of the Analysis Base can be explored to provide semantics which will help in achieving domain-based or keyword-based search performance, the creation of reference data and to enable reasoning. One of the best possible ways to achieve this is by building an ontological knowledge base of large-scale (Big) datasets, which includes the definition of a semantic model, the specification of domain knowledge, and the definition of links between different types of semantic knowledge.

### *Acknowledgments*

The authors would like to acknowledge the support of the European Union in funding this work via the neuGRID4You (N4U) project (grant agreement n. 283562, 2011-2014), with special thanks to the N4U consortium.